\documentclass{elsart}

\usepackage{amssymb}
\usepackage{epsf,colordvi,color,amsbsy}
\usepackage[dvips]{graphicx}
\begin{document}

\begin{frontmatter}

\title{On the Distinct Periodicities of Sunspot Counts in Flaring and Non-flaring Active Regions}

\author[]{Stefano Sello \corauthref{}}

\corauth[]{stefano.sello@enel.com}

\address{Mathematical and Physical Models, Enel Research, Pisa - Italy}

\begin{abstract}

In a recent work, Kilcik et al. (2017), have detected the temporal and periodic behavior of sunspot counts (SSC)
in flaring (i.e. C, M, or X class flares), and non-flaring active regions for the last two
solar cycles, covering the period: 1996 - 2016. The main results obtained are: 1) The temporal behavior of monthly means of daily total SSC in flaring and non-flaring active regions are different and these differences are also varying from cycle to cycle; 2) The periodicities detected in SSC of flaring and non-flaring active regions are quite different and these variations are also different from one cycle to another; the highest detected period in the flaring active regions is 113 days, while there are much higher periodicities (327, 312, and 256 days) in non-flaring regions. The detection of typical different periodicities in flaring and non-flaring regions can suggests both important differences and physical interpretation in the magneto-hydrodynamic behavior of the Sun. For this reason in the present paper we show a further periodicity analysis of the sunspot counts in flaring and in non-flaring active regions using the same data source of that used by the above cited authors and applying a powerful wavelet analysis tool which is particularly useful to detect multiscale features of complex unsteady and unevenly sampled time series. In order to futher support the differences and similarities found  in the time behavior of SSC in flaring and non-flaring regions, we also computed the behavior of the wavelet entropy, a proper time  function which allow us to measure the degree of complexity in the dynamics of the related time series.
 
 \end{abstract}
\end{frontmatter}

 \section{Introduction}

The present work was inspired by a recent paper by Kilcik et al. \cite{Kilcik17} where, for the first time, the authors have analyzed temporal and periodic behavior of sunspot counts (SSC) in flaring (C, M, or X class flares), and non-flaring active regions (AR) for the two last solar cycles (1996 - 2016). The idea is to distinguish the time behavior and the intrinsic periodicities in these two different kinds  of active regions in terms of magnetic complexity and thus of flares produced. The results obtained appear quite interesting: 1) The temporal behavior of monthly means of daily total SSC in flaring and non-flaring active regions are different and these differences are also varying from cycle to cycle; 2) The periodicities detected in SSC of flaring and non-flaring active regions are quite different and these variations are also different from one cycle to another; the highest detected period in the flaring active regions is 113 days, while there are much higher periodicities (327, 312, and 256 days) in non-flaring regions.
As noted by the above authors, high-energy solar activity is produced by the energy stored inside active region magnetic fields. Consequently flares often occur at sites in active regions overlying neutral lines where fields are strongly sheared and entangled. Further, there is a general trend for large active regions area  to
produce large flares and for more topologically complex regions to generate more numerous and
larger flares than other regions of similar size. Therefore it seems quite interesting to separate active regions in two main categories based on their flare productivity: if an active region or sunspot group produced a flare (C, M, and X class flares) during its natural evolution we count it as a flaring region, otherwise we count it as a non-flaring region.
The next step is to analyze the temporal and period behavior of the related sunspot counts
(SSC) in these regions: the hope as noted by the above authors is to "bring better understanding of the solar
cycle, the background physical phenomena, and consequently it may increase the knowledge about Sun-Earth interactions."
For these reasons it appears useful to further support and to investigate this kind of analysis, started by the above authors.
Here we use a powerful wavelet analysis method  which is particularly useful to detect multiscale features of complex unsteady and unevenly sampled time series and, in order to support the differences and/or similarities found  in the time behavior of SSC in flaring and non-flaring regions, we also computed the behavior of the wavelet entropy, a proper function  of time which allow us to measure the degree of dynamical complexity of the related time series, as in the solar activity time indices,  Sello, \cite{Sello00}, \cite{Sello03}.  
 
\section{Wavelet analysis and wavelet entropy}

Since the introduction of the wavelet transform general formalism by Grossmann and Morlet, 1984, in order to overcome the window limitations of the Gabor transform to non-stationary signals, this technique has been extensively applied in time-series analysis, including the study of solar and stellar activity cycles (Ochadlick,et al. 1993, Lawrence, et al. 1995, Oliver, et al. 1998). With a local decomposition of a multiscale signal, wavelet  analysis is able to properly detect time evolutions of the frequency distribution. This is particularly important when we consider intermittent and, more generally, non-stationary processes. More precisely, the continuous wavelet transform represents on optimal localized decomposition of a real, finite energy, time series: $x(t) \in L{^2}(\Re)$ as a function of both time, $t$, and frequency (scale), $a$, from a convolution integral:

\begin{equation}
(W_{\psi}x)(a,\tau) = \frac{1}{\sqrt{a}} \int_{-\infty}^{+\infty}
dt~x(t)\psi^* (\frac {t-\tau}{a})
\label{eq:wav1}
\end{equation}

where $\psi$ is called analysing wavelet if it verifies an admissibility condition:

\begin{equation}
c_{\psi} =  \int_{0}^{+\infty}
d \omega ~ \omega^{-1} {\arrowvert \hat{\psi}(\omega) \arrowvert}^2 < \infty
\label{eq:adm1}
\end{equation}

with:

\begin{equation}
\hat{\psi}(\omega)=\int_{-\infty}^{+\infty}
dt~ \psi(t) \mathrm{e}^{-i \omega t}
\label{eq:adm2}
\end{equation}

This last condition imposes: $\hat{\psi}(0)=0$, i.e. the wavelet has a zero mean. In eq.(\ref{eq:wav1}) $a,\tau \in {\Re}, (a \neq 0)$ are the scale and translation parameters, respectively (Daubechies, 1992, Mallat, 1998).
In fact, it follows from eq.(\ref{eq:wav1}) that the effectiveness of the wavelet analysis depends
on a suitable choice of the analyzing wavelet for the signal of interest. For our time-series application, where we are mainly interested to track the temporal evolution
of both the amplitude and phase of solar activity signals, we choose to use the family of complex analyzing wavelets
consisting of a plane wave modulated by a Gaussian, called Morlet wavelet (Torrence and Compo, 1998):

\begin{equation}
\psi(\eta) =  \pi ^{-1/4} \mathrm{e}^{i\omega_{0} \eta} \mathrm{e}^{-\eta^{2}/{2\sigma^2}}
\label{eq:wav4}
\end{equation}

where: $\eta={{t- \tau} \over {a}}$, and $\omega_{0}$ is a non-dimensional frequency. $\sigma$ is an adjustable parameter which can be determined in order to obtain the optimal wavelet resolution level both in time and frequency, for the characteristic time-scale of the original series (gapped wavelet, Soon, et al. 1999).
The limited frequency resolution imposes an half-power bandwidth of our wavelet given by: ${\Delta f \over f} \approx 0.12$. In the following computations we found as optimal general value: $\sigma \approx 1$. However, for specific cases where we need more frequency resolution, we have used a proper higher value.
The local wavelet spectrum at frequency $k$ and time $t$ and generally visualized by proper color contour maps is:

\begin{equation}
P(k,t) = {1 \over {2c_\psi k_0}} {\arrowvert W({k_0 \over k},t) \arrowvert} ^2,~k \geq 0
\label{eq:ent3}
\end{equation}

In eq.(\ref{eq:ent3}) $k_0$ is the peak frequency of the analyzing wavelet $\psi$ (Torrence and Compo, 1998).

Following Foster (1996), here we consider an extension of the above wavelet 
formalism in order to correct handle irregularly sampled time series. The wavelet 
transform is viewed, in the time domain, as a suitable weighted projection onto three trial functions
giving the Weighted Wavelet Z transform and the Weighted Wavelet Amplitudes. For
all the mathematical details of this formulation and its applications we refer
to Foster paper \cite{Foster96}. One of the advantages of this approach is the computation of eq.(5) in the time-domain, allowing us to avoid some limitations typical of the
computation in the Fourier domain, as shown in Torrence and Compo (1998).

Many applications of the wavelet analysis suffered from an apparent lack of 
quantititative evaluations especially by the use of arbitrary normalization and the 
lack of statistical significance test in order to estimate the reliability of results.
 Here we used power spectrum normalization and significance levels following
 the approach suggested by Torrence et al. \cite{Tor98}. We first assume an appropriate
background spectrum and we suppose that different realizations of the considered physical process will be randomly distributed about this expected background
spectrum. The actual spectrum is compared against this random distribution. In the
present work we assumed as background spectrum a red noise spectrum modeled
through a univariate lag-1 autoregressive process:

\begin{equation}
x_n =  \alpha_{x_{n-1}}+z_n
\end{equation}
where: $z_n$ is derived from a Gaussian white noise and $\alpha$ is the lag-1 autocorrelation here estimated by:

\begin{equation}
\alpha={\alpha_1+\sqrt{\alpha_2} \over{2}}
\end{equation}
where: $\alpha_1$ and $\alpha_2$ are respectively the lag-1 and lag-2 autocorrelations of the considered time
series. The discrete normalized Fourier power spectrum of this red noise is:

\begin{equation}
P_k={ {1- \alpha ^2} \over {1+ \alpha ^2-2 \alpha \cos({2 \pi k \over N})}} 
\end{equation}
and the following null hypothesis is defined for the wavelet power spectrum: we
assume the red noise spectrum as the mean power spectrum of the time series; if
a peak in the wavelet power spectrum is significantly
above this background spectrum (typically at $95\%$ or $99\%$ confidence level), then it can be considered to be a real feature of the time series, Torrence and Compo, \cite{Tor98}.

From the local wavelet power spectrum obtained from eq.(5), it is
possible to derive various time dependent functionals. An interesting
example is related to the measure of the "disorder" level contained in the signal
well quantified by the \emph{wavelet entropy} introduced by Quian Quiroga
and co-workers \cite{Quir99}. The concept of thermodynamic entropy is well
known in physics as a measure of the system disorder. A previous measure 
of entropy has been introduced from the Fourier power spectrum applied to
Hamiltonian systems, called spectral entropy, by Powell and Percival \cite{Powell79}.
Using the extended technique of the wavelet formalism, it is possible to
define a discrete or continuous wavelet entropy as a function of time, Quian Quiroga et al., \cite{Quir99}, Sello, \cite{Sello00}:

\begin{equation}
WS(t)=-\int_{0}^{+\infty}{dk~p_{t,k}log_2(p_{t,k})}
\label{eq:ent}
\end{equation}
 
where:

\begin{equation}
p_{t,k}={P(k,t) \over {\int_{0}^{+\infty}{dk~P(k,t)}}}
\label{eq:ent2}
\end{equation}

is the power probability distribution for each scale level $k$ at time $t$. Of course, in many real applications, where we know the involved functions only in a given set of discrete values, the integrals in eqs.(9),(10) become finite sums.

From  eq.(\ref{eq:ent}), it follows that the wavelet entropy is minimum
when the signal represents an ordered activity characterized by a narrow frequency 
distribution i.e. there are few relevant frequencies involved, whereas the entropy is high when the signal contains 
a broad spectrum of frequency distribution, i.e there are many relevant frequencies involved. This last feature is a common sign
related to some kind of dynamical complex behavior.
Here we are mainly interested in distinguishing between ordered or regular
processes, and more complex dynamical behaviors involved in solar activity.
In fact, a genuine stochastic process and a deterministic chaotic dynamics 
are both characterized by a broad-band spectrum, with many relevant  interacting
frequencies. However, as follows from definition, eqs. (9), (10), the wavelet entropy alone is not able to distinguish the above
two different cases of complex dynamics.
Previous analyses on different solar activity indices, have 
 illustrated clearly the usefulness of
the wavelet entropy to follows the time evolution of the complexity level of solar cycle activity, Sello, \cite{Sello00}, \cite{Sello03}.

\section{SSC data and results}

In this work, we have considered the same data set used by Kilcik et al. \cite{Kilcik17} to investigate time variations and periodicities of sunspot counts
depending on the flare production of active regions. The group classification and the flare
data are only available since August 1996 and thus the analyzed time interval
covers a limited time interval from solar cycle 23 (1996 through 2008) and the ascending, maximum and early descending
phases of current solar cycle 24 (2009 through 2016). The raw data are taken from
the Space Weather Prediction Center (SWPC), including all
X-Ray flares and active regions information. Then the daily total sunspot counts and related monthly mean values for flaring
and non-flaring ARs were calculated. For our analysis it is more useful to consider the 12-points monthly smoothed sunspot numbers, which contain less high frequency fluctuations.
In Figure 1 we compare the behavior of flaring and non-flaring aunspot counts with the revised total monthly smoothed sunspot numbers from SIDC, (WDC-SILSO
\cite{SIDC}). Note that this new version is $40\%$ - $70\%$ larger than the previous older version, but here we are mainly interested to the phase of sunspot numbers. The figure clearly shows that there is a synchronous information for all the sunspot counts.

\begin{figure}
\resizebox{\hsize}{!}{\includegraphics{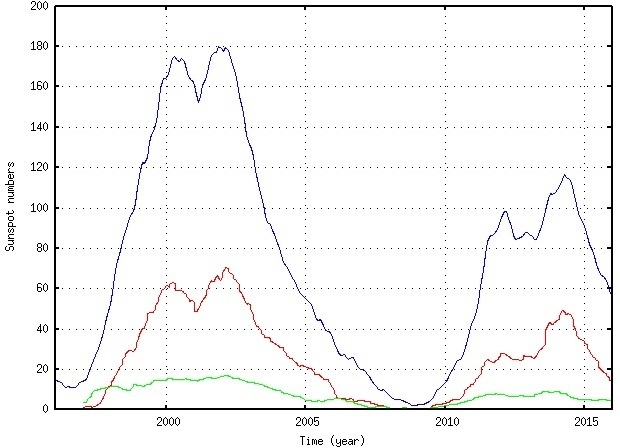}}
 \caption{Comparison between flaring (red) and non-flaring (green) sunspot counts with the total monthly smooted sunspot numbers SIDC (blue). Note that all the sunspot counts are well synchronous or in phase.} \
 \label{lab1}
\end{figure}
  
The analysis of flaring and non-flaring sunspot counts are here performed using the methodology described in Section 2. We note that  Kilcik et al. \cite{Kilcik17}, 
 in order to reveal periodic variations in the data, have used two period analysis methods: Multitaper Method (MTM) and Morlet Wavelet Analysis, following the main procedures as described in Torrence and Compo, \cite{Tor98}.

In Figure 2 we show our results of the wavelet analysis of flaring sunspot counts for both the 23rd and 24th solar cycles. Note that  Kilcik et al. \cite{Kilcik17} show their results separating the analyses for solar cycle 23 and for solar cycle 24 (Figures 2-3 and Figures 4-5 in the cited paper).

\begin{figure}
\resizebox{\hsize}{!}{\includegraphics{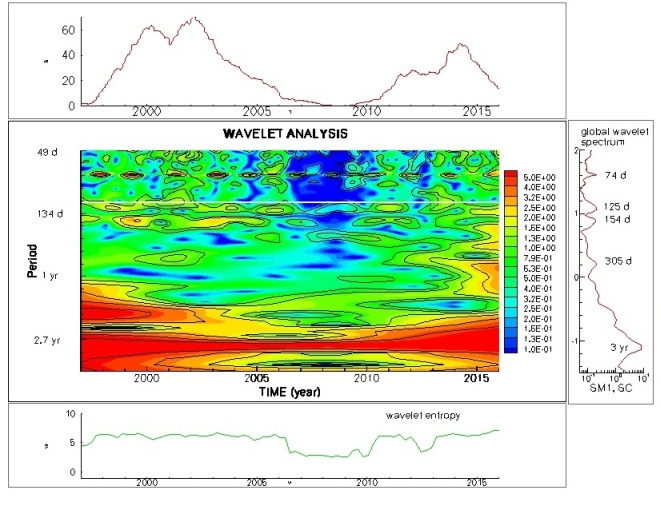}}
 \caption{Wavelet analysis of flaring sunspot count regions. White line on the wavelet map corresponds to a 113-days period. See text for a complete explanation.} \
 \label{lab2}
\end{figure}

Referring to Figure 2, the upper panel shows the original time series in its natural units (red line); the central panel shows the amplitudes of the wavelet local power spectrum in terms of a multi-colour contour map: red corresponds to the strongest energetic contributions to the power spectrum, while blue represents the weakest components. Horizontal time axis corresponds to the axis of the upper time series and  the vertical scale or frequency axis is, for convenience, expressed in log
values of cycle per year$^{-1}$ where here it is shown the related period ($>49$ days). The right panel shows the mean wavelet spectrum (an averaged and weighted Fourier spectrum) obtained by a selected time integration of the local map in logarithmic scale. We have labeled the principal peaks detected. The $95\%$ significance of the local power map was tested using an adjustable red-noise autoregressive lag-1 background spectrum and here localized by black innermost contour lines. Finally, in the bottom panel the green  line is the computed wavelet entropy, properly scaled in order to clearly demonstrate  its time behavior when compared with the original time series.
As clearly shown in Figure 2, the principal periods found are: $74$ days; $125$ days; $154$ days; $305$ days and $3$ years. As we can see from the wavelet map there are some differences between the behavior of the two cycles but also many similarities, except from the more strong presence of the $154$ days periodocity during the 23rd cycle and the more strong presence of $305$ days periodicity during the 24th cycle. Note the recurrent presence of a local peak in the wavelet map at the $74$ days periodicity in both cycles that appears as a beating frequency. In fact, using the gapped wavelet with adjustable $\sigma$ parameter, a higher value ($\sigma=8$) reveals that the above period is splitted in two narrow periods: $70.0$ and $75.7$ days. Wavelet entropy behavior shows a quite persistent level during cycle 23rd with an expected lower level during the minimum phase of the cycle; whereas for cycle 24th there is a more variable level with a deep reduction (similar to the minimum phase level) after the first peak before reaching the Gnevyshev's gap phase where, in general, we observe an increase of dynamical complexity probably related to an enhanced number of low intensity and comparable energy processes involved, which in turn produces a spread in the frequency 
distribution and a higher wavelet entropy (see: Sello, \cite{Sello03}). The same reduction is visible during cycle 23rd but less marked. The early descending phase of cycle 24th is characterized by an increase of wavelet entropy i.e. involving more complex multiscale processes.

For a first comparison, we recall here the main results obtained by Kilcik et al. \cite{Kilcik17} for flaring susnspot counts: i) for cycle 23rd, the largest detected period in the flaring active regions is 113 days, while there are much higher periodicities in non-flaring active regions. The common presence of similarity between the periodicities of the two data sets is the existence of solar rotation periods (27-35 days) but here these short periods are not considered, ii) for cycle 24th, a 45 day periodicity appear only in flaring active regions of both solar cycles 23rd and 24th.

In Figure 3  we show our results of the wavelet analysis of non-flaring sunspot counts for both the 23rd and 24th solar cycles.
As clearly shown in Figure 3, the principal periods found are: $74$ days (splitted in $70.0$ and $75.7$ days); $125$ days; $154$ days; $221-308$ days; $1.33$ years, $1.97$ years and $2.72$ years. As we can see from the wavelet map the first three periods are common with the flaring counts but there are differences for the  largest periodicities where for non-flaring counts appears more periods higher than one year. One again there are differences and similarities between the two data sets.
 Wavelet entropy behavior shows here a clearly different behavior: the variability is very high and there exist more longer periods with a very low level of entropy  indicating that for non-flaring regions there are longer periods of quite ordered dynamical states here not strongly perturbed by energetic flares.
 
For a first comparison, we recall here the main results obtained by Kilcik et al. \cite{Kilcik17} for non-flaring susnspot counts: i) for cycle 23rd, there are much higher periodicities: 327, 312, and 256 days;  ii) for cycle 24th, we detect a 72 days periodicity and some larger periodicities, about 250 and 600 days, appear in the wavelet scalograms of non-flaring active regions.The early descending phase of cycle 24th is characterized by a clear decrease of wavelet entropy i.e. involving few complex multiscale processes.

\begin{figure}
\resizebox{\hsize}{!}{\includegraphics{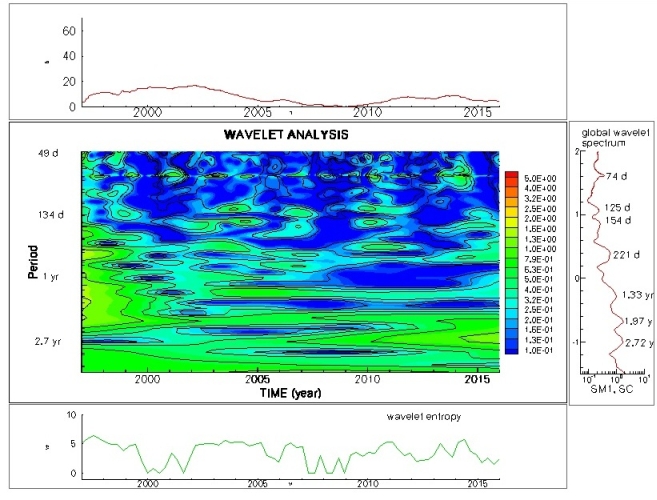}}
 \caption{Wavelet analysis of non-flaring sunspot count regions. See text for a complete explanation.} \
 \label{lab3}
\end{figure}

In the next section we discuss the detected periodicities with the different analyses.

\section{Comments and discussion}

Time behavior and periodic variations observed in sunspot counts of flaring and non-flaring active regions for cycles 23rd and 24th, show important differences but also similarities: in the present analysis we found more similarities than those detected by Kilcik et al. \cite{Kilcik17}. In particular, $74$ (splitted in $70.0$ and $75.7$) days; $125$ days; $154$ days are detected in both the data with different amplitudes but with very similar time/frequency characteristics. When we consider larger periodicities we found important differences. Moreover, as expected, the data differ in the wavelet entropy behavior due to the presence or absence of high energetic processes, such as energetic flares with involved complex multiscale dynamics.
As well discussed by Kilcik et al. \cite{Kilcik17} many previous studies were focused on the interpretation of temporal variation and involved periodicities of many different solar activity indices  or indicators (e.g. SSCs, SSAs, Solar X-Ray flares, etc.). It is quite interesting to note that our analysis detected a periodicity at the well known Rieger-type periodicity, around $154-158$ days, discovered studying $\gamma$-ray and X-ray flare data from Solar Maximum Mission (SMM) in solar cycle 21, and linked to energetic solar flares, Rieger et al. \cite{Rieger84}. Further, a near $153$-day period was found in other different solar activity indices: in H-$\alpha$ flare importance, hard X-ray peak rate, in flare index, 10.7–cm radio peak flux, etc. (For a detailed  list of related references see: Kilcik et al. \cite{Kilcik17}). Also the near $74$ days periodicity was found by many authors using various solar activity indicators. This wide and differentiated spectrum of solar activity may be interpreted in terms of different deterministic and stochastic magnetic process components which act with different amplitudes and phases during a given solar cycle. For this reasons we expect a variability in time behavior from cycle to cycle but also some common features related to the same source of solar magnetic dynamo. This is confirmed in our wavelet analysis of flaring and non-flaring sunspot counts. Due to  different spatial scales involved in magnetic regeneration and evolution during a solar cycle, we can distinguish roughly two main groups of periodicities, as suggested by Kilcik et al. \cite{Kilcik17}: the first group of periods around 1-2 years may be associated with the evolution of large-scale magnetic fields, and these periods have been detected in our analysis. The second group of shorter periods, less than 300 days, are mainly related to more local magnetic processes inside active regions and these are clearly visible with some differences both in solar cycles 23rd and 24th near maxima and for both flaring and non-flaring data. In fact, the more distintive feature found between the flaring and non-flaring data is the behavior of the wavelet entropy and the emergence, in the last group, of more periodicities greater that 1 year and this can be a consequence of a more important presence of the more regular and stable large-scale fields which dominate the underlying dynamics. The 113-days periodicity found by Kilcik et al. \cite{Kilcik17}) only in cycle 23rd and only for flaring data, is in fact present in our wavelet analysis but only in two very localized weak events near the mddle of the Gnevyshev's gaps: 2001.3 for cycle 23rd and 2013.6 for cycle 24th in flaring data (see white line on the wavelet map in Fig.2) (in non-flaring data it is only visible in cycle 24th), suggesting that this periodicity may be linked to a reorganization of the large-scale magnetic field probably connected to the polar heliomagnetic field reversal rather than a characteristic of solar flares as suggested by Kilcik, \cite{Kilcik17}; on the other hand the longer, more persistent and common 125 days period is, as expected, more prominent in flaring group and thus it appears as a better candidate.
Of course, in order to give more conclusive and statistically robust results, we need more data spanning a longer time interval covering more solar cycles and this will be the goal of future analyses.

\section{Conclusions}

Starting from a recent paper by Kilcik et al. \cite{Kilcik17} where, for the first time, the authors have analyzed temporal and periodic behavior of sunspot counts (SSC) in flaring (C, M, or X class flares), and non-flaring active regions (AR) for the two last solar cycles (1996 - 2016), in this work we performed a similar analysis using a proper version of  wavelet analysis tool which is particularly useful to detect multiscale features of complex unsteady and unevenly sampled time series. In order to futher support the differences and similarities found  in the time behavior of SSC in flaring and non-flaring regions, we also computed the behavior of the wavelet entropy, a proper time function which allow us to measure the degree of complexity in the dynamics of the related time series.
Time behavior and periodic variations observed in the two group of data, show important differences but also similarities: in the present analysis we found more similarities than those detected by Kilcik et al. \cite{Kilcik17}. In particular, $74$ (splitted in $70.0$ and $75.7$) days; $125$ days; $154$ days are detected in both the data with different amplitudes but with very similar time/frequency characteristics. When we consider larger periodicities we found important differences. Moreover, as expected, the data differ in the wavelet entropy behavior due to the presence or absence of high energetic processes, such as energetic flares with involved complex multiscale dynamics.
The more distintive periodic feature found between the flaring and non-flaring data is the emergence, in the last group, of more periodicities greater that 1 year and this can be a consequence of a more important presence of the quite regular and stable large-scale magnetic fields which dominate the underlying dynamics.
For more conclusive and statistically robust results, we need more data spanning a longer time interval and covering more solar cycles. However, these analyses may help us to better understand and to better interpret the dynamics of magnetic processes underlying the solar dynamo mechanism which drives the solar activity both on large-scale and on local scale, where we observe the most energetic solar events such as the solar flares.

\end{document}